%% file: main.tex
\documentclass[runningheads]{llncs}
\usepackage[T1]{fontenc}
\usepackage{amssymb}
\usepackage{amsmath}
\usepackage{amsthm}
\usepackage{enumerate}
\usepackage[inline]{enumitem}
\usepackage{xcolor}
\usepackage{marvosym}
\usepackage{lineno}

\newtheorem{mechanism}{Mechanism}

\DeclareMathOperator{\dist}{dist}
\DeclareMathOperator{\cost}{cost}
\DeclareMathOperator{\SC}{SC}
\DeclareMathOperator{\OPT}{OPT}
\DeclareMathOperator{\Dist}{Dist}

\begin{document}
\title{On the Distortion of Multi-winner Election Using Single-Candidate Ballots}
%
%
\author{Gennaro Auricchio\inst{1} \and Zeyu Ren\inst{2} \and Zihe Wang\textsuperscript{(\Letter)}\thanks{Zihe Wang was supported by National Natural Science Foundation of China (Grant No. 62172422) and by fund for building world-class universities (disciplines) of Renmin University of China.}\inst{2} \and Jie Zhang\thanks{Jie Zhang was partially supported by a Leverhulme Trust Research Project Grant (2021 -- 2024) and the EPSRC grant (EP/W014912/1).}\inst{3}}

\institute{
University of Padua, Via 8 Febbraio, 2, Padova, 35122, Veneto, Italy\\
\email{gennaro.auricchio@unipd.it}
\and
Renmin University of China, 59 Zhongguancun Street, Beijing, 100872, Beijing, China\\
\email{\{zeyuren, wang.zihe\}@ruc.edu.cn}
\and
University of Bath, Claverton Down, Bath, BA2 7AY, Somerset, United Kingdom
\email{jz2558@bath.ac.uk}
}
%
\authorrunning{G. Auricchio et al.}
%
%
\maketitle              
\begin{abstract}
In this paper, we study the distortion bounds for voting mechanisms in multi-winner elections in general metric spaces.
Our study pertains to the case in which each voter only reports her favorite candidate amongst $m$ possible choices. 
Given that candidates' locations are undisclosed to the mechanism, the mechanism has to form a $w-$winner committee based solely on the number of votes received by candidates.
We establish distortion bounds for both truthful and non-truthful mechanisms. Our research highlights the significance of the $\sigma$ parameter, which represents the ratio between maximum and minimum distances among all candidate pairs. 
We show that the distortion is linear in $\sigma$.
First, we demonstrate that all mechanisms possess a distortion greater than $1+\frac{w-1}{w+1}(\sigma-1)$.
To give an upper bound, we study the Single Non-Transferable Vote (SNTV) mechanism, whose distortion is at most $1+2\sigma$.
Second, we retrieve the upper bounds for strategyproof mechanisms.
In particular, we infer an upper bound by examining the Random Sequential Dictator mechanism that achieves a distortion less than $1+4\sigma$ when $w=2$.
\end{abstract}
\input{introduction}
\input{setting}
\input{non-sp}
\input{sp}

\section{Conclusion and Future Work}
In this paper, we studied the problem of multi-winner voting using single-candidate ballots. Our study examined mechanisms in non-strategyproof and strategyproof settings, where we identified the lower and upper bounds of distortion.
Overall, our research contributes to understanding multi-winner voting problems, offering insights into distortion bounds and the significance of the parameter $\sigma$.

There are still several unresolved questions that merit further exploration. First, an intriguing alternative is to explore mechanisms where each voter is asked to provide their top-$k$ choices or even a rank of all candidates rather than simply indicating their favorite. Secondly, a compelling question arises: can our approaches be adapted to address the max cost objective? If so, to what extent would this alteration impact the distortion bounds? These inquiries present formidable challenges and warrant in-depth investigation.

\bibliographystyle{splncs04}
\bibliography{ref}

\clearpage
\appendix
\section*{APPENDIX}
\setcounter{section}{0}
\input{appendix.tex}

\end{document}

%% file: introduction.tex
\section{Introduction}
Algorithmic mechanism design operates at the intersection of computer science, game theory, and economics. 
The field is concerned with designing mechanisms that achieve desired objectives while possessing appealing social properties, such as strategyproofness, which ensures that agents reveal their preferences.
An important research area within algorithmic mechanism design is determining how, given the voters' preferences, a mechanism can select a committee of winners from a pool of candidates.
A standard metric for quantifying the quality of a mechanism is the distortion, proposed by \cite{procaccia2006distortion}, which is the worst-case ratio between the social objective attained by the mechanism and the optimum.
Determining the trade-offs between attaining an optimal outcome and other desirable social properties is a key problem in algorithmic mechanism design.
The bounds on the distortion achievable by an election mechanism depend on the information disclosed by voters to the mechanism. 
The study by \cite{DBLP:conf/sigecom/FeldmanFG16} examines three types of mechanism in the facility location-based voting problem:
\begin{enumerate*}[label=(\roman*)]
    \item \textit{voting mechanisms}, where agents vote for their preferred candidate;
    \item \textit{ranking mechanisms}, where agents report their ordinal preferences for candidates; and 
    \item \textit{location mechanisms}, where agents reveal their position in a political spectrum, i.e. their positions or view on different topics.
\end{enumerate*} 
The more information agents share, the more resources mechanisms have to reduce the distortion.
In this paper, we focus on the first category of mechanisms, that is voting mechanisms.
This class of mechanisms is appealing for many reasons.
First, voters may find it impossible to provide rankings for different candidates, primarily for cognitive reasons. 
In addition, agents' willingness to disclose their location can vary due to factors such as privacy concerns, fear of discrimination, and mistrust in how their information is used. 
Within this framework, we explore the distortion of multiwinner problems in the metric space; thus the mechanism collects only the first preference of voters and returns a committee of $w$ winners, where $w>1$.
Unfortunately, a simple example shows that the distortion is unbounded in this case: Let $y_1,y_2$ and $y_3$ be three candidates.
Let us assume that there are three voters, each one of them shares the position with one of the candidates, and that the mechanism is asked to return a committee of two winners. Without loss of generality, we assume that the mechanism assigns the committee $\{y_1,y_2\}$ a positive probability. 
However, if the candidate profile is such that $y_1, y_2$ are located at the same point and $y_3$ is far away from them. Then the expected social cost achieved by the mechanism is positive and arbitrarily large, while the optimal cost is 0.
%
In particular, the distortion is unbounded in this case. 
%
%
This is due to the fact that the mechanism has only access to the number of votes each candidate gets, but not to the candidates' locations.
To overcome this issue, we introduce a parameter that represents the ratio between the maximum and minimum distance between candidate pairs, namely $\sigma\triangleq\frac{d_{\max}}{d_{\min}}$.
%
%
The parameter $\sigma$ allows us to better characterize both the lower and upper distortion bounds of the problem, allowing us to address the voting problem for any generic metric space.
We consider both strategyproof and non-strategyproof mechanisms in our study.

\subsection{Our Contribution and Technique Overview}
In this paper, we study the distortion bounds of voting mechanisms for multi-winner elections in a general metric space.
In particular, we assume that each voter submits only their favorite candidate to the mechanism. 
Based on this information, the mechanism must form a committee of $w$ winners.
%
%
In this framework, we first study the lower and upper bounds for general voting mechanisms.
We show that both distortion bounds are linear in $\sigma$, which is the ratio between the maximum and minimum distance between any pair of candidates.
In particular, we show that the distortion of any mechanism is lower bounded by $1+\frac{w-1}{w+1}(\sigma-1)$ and study the Single Non-Transferable Vote (SNTV) mechanism, a greedy mechanism that returns the $w$ candidates who receive the most votes and arbitrarily breaks ties.
We show that the SNTV mechanism has a distortion less than $1+2\sigma$.
We then study the distortion bounds for strategyproof mechanisms.
We first focus on randomized mechanisms and then consider deterministic mechanisms.
%
We show that every truthful mechanism is \textit{independent of irrelevant candidates}, which ensures that the probability of any committee $C$ remains unaffected by the votes received by candidates outside the committee. 
%
%
%
%
We then introduce the Random Sequential Dictator mechanism, which achieves an upper bound of at most $1+4\sigma$ when $w=2$.
Lastly, we study deterministic truthful mechanisms. 
We establish that no anonymous deterministic strategyproof mechanism can achieve finite distortion. 
We show, however, that if we relax the anonymity condition and denote with $n$ the number of voters, the Sequential Dictator mechanism achieves a distortion bound of $2(n-w)\sigma+1$.

The most significant findings of our research are \begin{enumerate*}[label=(\roman*)]
\item the role played by the ratio between the maximum and minimum distances of any pair of candidates, that is $\sigma$.
Introducing the $\sigma$ ratio allows for a more fine-grained analysis of distortion in multiwinner voting problems. 
\item a property of truthful mechanisms.
Indeed, we show that every truthful mechanism is independent of irrelevant candidates and use this property to retrieve the lower bound of strategyproof mechanisms.
\end{enumerate*}

\subsection{Related Work}
To evaluate the performance of voting rules, Procaccia and Rosenschein \cite{procaccia2006distortion} introduced the notion of distortion into the normalized social choice setting. Later, Caragiannis and Procaccia \cite{caragiannis2011voting} employed this notion and analyzed the Plurality Rule.  Boutilier et al. \cite{boutilier2015optimal} studied the distortion of randomized rules and presented a simple rule with lower distortion. Caragiannis et al. \cite{caragiannis2017subset} extended the framework to select a subset of alternatives.

Anshelevich et al. \cite{anshelevich2015approximating} first initiated the study of this problem in a metric space.  Following their study, several papers analysed the distortion of many different rules: Skowron and Elkind \cite{skowron2017social} studied the class of scoring rules, Goel
et al. \cite{goel2017metric} and Kempe \cite{kempe2020analysis} studied the Ranked Pairs Rule,  Munagala and Wang \cite{munagala2019improved} introduced a weighted tournament rule and improved the distortion , while Gkatzelis et al. \cite{gkatzelis2020resolving} designed the plurality matching voting rule and proved that the optimal deterministic algorithm has distortion $3$.  Kizilkaya and Kempe proposed an extremely simple voting rule which achieves the same optimal distortion of 3.
As for randomized mechanisms, Anshelevich and Post \cite{anshelevich2017randomized} showed that Random Dictatorship has distortion $3-\frac2n$.
Kempe \cite{kempe2020communication} and Gkatzelis et al. \cite{gkatzelis2020resolving} improved on this result by providing a mechanism whose distortion is $3-\frac2m$.
Charikar et al. \cite{charikar2024breaking} further improved the results by showing that that randomization over simple rules can achieve distortion less than $2.753$, making it closer to the known lower bound of 2.1126, found by Charikar and Ramakrishnan \cite{charikar2022metric}. Pulyassary and Swamy \cite{pulyassary2021randomized} independently showed a lower bound of 2.0631.
Lastly, it is important to highlight the findings of \cite{gross2017vote}, who investigated the distortion of a randomized mechanism known as 2-Agree. This mechanism operates by sequentially querying random voters for their top choices until a consensus is reached among at least two voters.
Building on this line of research, Fain et al. \cite{fain2019random} examined mechanisms with constant sample complexity and introduced the Random Oligarchy mechanism.
For other distortion results, please refer to the survey by Anshelevich et al. \cite{anshelevich2021distortion}.
\emph{Multi-winner setting.}  
Goel et al. \cite{goel2018relating} characterized the distortion of selecting a committee by repeatedly applying single-winner voting rules. 
Chen et al. \cite{chen2020favorite} shifted their focus to the \textit{single-loser} setting, which is $w=m-1$. In this setting, the committee is formed by eliminating the least popular candidate. When considering the social cost objective, they demonstrate a tight distortion bound of 3 for deterministic mechanisms and $3-2/m$ for randomized mechanisms where $m$ is the number of candidates. While this paper mentioned is relevant to our work, a significant difference lies in our primary focus on general multi-winner elections.
Caragiannis et al. \cite{caragiannis2022metric} studied ranking mechanisms, where agents report their ordinal preferences for candidates. They showed that with all ordinal preferences for candidates, the distortion is asymptotically linear in the number of agents when $w=2$, and the distortion is unbounded when $w>2$. We studied voting mechanisms where agents vote for their preferred candidate. Our conclusion is different because we consider mechanisms using different information, and we represent the distortion in terms of $\sigma$. When $w=2$, we show a lower bound that is linear in $\sigma$. Since $\sigma$ can be arbitrarily large, then our result implies that the distortion is unbounded for mechanisms using single-candidate ballots. The result can be easily generalized for $w>2$. Aziz et al. \cite{aziz2017justified}, Kalayci et al. \cite{kalayci2024proportional}, and Peters and Skowron \cite{peters2020proportionality} studied the proportional representation. Please refer to \cite{faliszewski2017multiwinner} for other interesting topics in the multi-winner setting. 
%

%% file: setting.tex
\section{Preliminaries}
Let $\Omega = (S, d)$ be a metric space.
An election in $\Omega$ consists of $n$ voters, which we denote with $N=\{1,\dots,n\}$, and $m$  candidates, which we denote with $M=\{1,\dots,m\}$.
All candidates and voters are points in $S$, thus we denote with $x_i\in S$ the positions of voters and with $y_j\in S$ the positions of candidates.
We denote in bold letters the vector containing all the positions of the agents and all the positions of the candidates, that is $\mathbf{x}=(x_1,\dots,x_n)$ and $\mathbf{y}=(y_1,\dots,y_m)$.
For \emph{single-candidate ballots}, each voter $i$ votes for one candidate, which we denote with $a_i$.
The vote $a_i$ of voter $i$ is also referred to as its \textit{action}. 
The distance between a voter $i$ and her action $a_i$ is $d(x_i,a_i)$.
We denote the action profile by $\mathbf{a}=(a_1,\dots,a_n)$, a vector containing all the actions of voters.
An action profile $\mathbf{a}$ is \textit{consistent} with $\mathbf{x}$ if $a_i \in \arg\min_{y \in M} d(x_i,y)$ for every $i\in\{1,\dots,n\}$.
Given a voter at $x_i$ and a committee $C\subseteq M$, we define the cost of agent $i$ as her distance from $C$, that is
\[\cost(x_i,C) = d(x_i,C) \triangleq \min_{y_k\in C} d(x_i,y_k).\]
Given $w\in\mathbb{N}$, we denote with $\mathcal{C}_w$ the set of all committees of $w$ different candidates.
The problem of \emph{multi-winner voting} for single-candidate ballots consists in electing a committee $C^*\subseteq M$ of $w$ winners that minimizes the sum of the voters' costs i.e.
\[C^*\in\arg\min_{C\in \mathcal{C}_w}\sum_{i\in[n]}\cost(x_i,C).\]
The sum of the voters' costs is also known as the Social Cost (SC) of the committee $C$, thus, henceforth, we set $\SC(\mathbf{x}, C)\triangleq\sum_{i\in[n]}\cost(x_i,C)$.
An election in the social choice problem under consideration is a tuple $\Gamma=(\Omega, M, \mathbf{a}, w)$. For simplicity, we may refer to an election $\Gamma$ as its action profile $\mathbf{a}$. 

A mechanism $f$ takes an election $\Gamma$ as input and outputs a committee $C\subseteq M$ of $w$ winners.
A mechanism is \textit{deterministic} if, for each election, it outputs one committee.
A mechanism is \textit{randomized} if it outputs a probability distribution over committees in $\mathcal{C}_w$. The probability of the committee $C$ being elected is $p_C(\mathbf{a})$. 
Moreover, a mechanism $f$ is \textit{anonymous} if the output of the mechanism does not depend on the identities of the voters but only on the voters' aggregated information, that is the number of votes each candidate receives.
For the sake of simplicity, given an anonymous mechanism $f$ and an action profile $\mathbf{a}$, we denote the probability of a committee $C$ being elected with $p_C(\mathbf a) = p_C(n_1,\dots,n_m)$, where $(n_1,\cdots,n_m)$ is the $m$-tuple containing the number of votes of each candidate given a voters' action profile $\mathbf{a}$.
When voters are self-interested, we assume that the position of every agent is their own private information.
In this case, voters may act strategically if this lowers the cost.
A mechanism $f$ is \textit{truthful} (or \textit{strategyproof}) if no agent is able to lower its cost by misreporting their action, that is, for every $i\in N$ and every action $a_i'$, we have that  $\cost(x_i,f(a_i,\mathbf{a}_{-i})) \le \cost(x_i,f(a_i',\mathbf{a}_{-i}))$ where $a_i$ is an action consistent with the position of voter $i$ and $\mathbf{a}_{-i}$ denotes the actions of the other voters.
To evaluate the performance of a mechanism $f$, we consider its \textit{distortion}.
For a fixed election $\Gamma$ and the action profile $\mathbf{a}$, the distortion of $f$ over the election $\Gamma$ is defined as the worst-case ratio between the expected SC returned by $f$ and the optimal SC over all the agents' positions that are consistent with the action profile $\mathbf{a}$, that is 
\begin{equation*}
    \dist(f,\Gamma)=\sup_{\mathbf{x}\in \chi(\Gamma)}\frac{\mathbb{E}[\SC(\mathbf{x}, f(\mathbf{a}))]}{\OPT(\mathbf{x})},
\end{equation*}
where  $\chi(\Gamma)$ is the set of location profiles consistent with $\Gamma$ and $\OPT(\mathbf{x})$ is the SC of the optimal solution.
Finally, the distortion of a mechanism $f$ is then $\Dist(f) = \sup_{\Gamma} \dist(f, \Gamma)$, which is the worst case in all elections. For deterministic mechanisms $f$, the distortion is defined similarly.
In our study, we show that the distortion is strongly related to the ratio between the maximum and minimum distances among any two candidates. We denote this ratio as $\sigma$. Formally, we express the ratio as $\sigma=d_{\max}/d_{\min}$, where
$d_{\max}=\max_{(y_k,y_l)\in M^2}d(y_k,y_l)$ and $d_{\min} = \min_{(y_k,y_l) \in M^2}d(y_k,y_l)$. 

%% file: non-sp.tex
\section{Distortion Without Strategyproofness}
\label{sec:non-sp}
In this section, we study the distortion bounds of mechanisms that are not necessarily strategyproof.
We start our analysis from the lower bound and show that no randomized voting mechanism can achieve an approximation ratio that is lower than $1+\frac{w-1}{w+1}(\sigma-1)$.
\begin{theorem}\label{thm:lb_nonsp}
It is impossible for any randomized mechanism to achieve a distortion smaller than $1+\frac{w-1}{w+1}(\sigma-1)$.
\end{theorem}
In order to prove the lower bound, we first introduce a candidate profile.
\begin{definition}[Candidate Profile I]
    Let us consider $m$ candidates in an $m$-dimensional Euclidean space. The coordinate of the locations of $y_i(1\le i\le m-2)$ has only one dimension that is non-zero.
    Given $r\in\mathbb{R}$, we denote the locations of $y_i(1\le i\le m-2)$ as $y_1=(r,0,\dots,0),y_2=(0,r,\dots,0),\dots, y_{m-2}=(0,\dots,0,r,0,0)$. The locations of $y_{m-1}$ and $y_m$ are $y_{m-1}=(0,\dots,0,\sqrt{r^2-1},1)$ and $y_m=(0,\dots,0,\sqrt{r^2-1},-1)$.
    When $r$ is large enough, the distance between $y_{m-1}$ and $y_m$ is $d_{\min}$. Any other distance between two candidates is $d_{\max}$.
\end{definition}
Using the candidate profile, we are ready to prove Theorem~\ref{thm:lb_nonsp}. 

\begin{proof}
Suppose that $n$ voters are co-located at $w+1$ candidates, that is, each candidate gets $\frac{n}{w+1}$ votes. The $w+1$ candidates are $y_{m-w},\cdots,y_{m-1}$ and $y_m$. 

First, the committee $\{y_{m-w},\cdots,y_{m-1}\}$ or $\{y_{m-w},\cdots,y_{m-2},y_m\}$ is the optimal. Thus, the optimal social cost is $\frac{n}{w+1}\cdot d_{\min}$. 

Then, we consider the social cost achieved by the mechanism. Let $p_t$ denote the probability of selecting $w$ candidates from $w+1$ candidates that receive votes. Due to the mechanism designer does not know the exact locations of candidates, she should treat $w+1$ potential committees equally. It indicates that the probability of any committee is $\frac{p_t}{w+1}$. Otherwise, we can do a permutation over all the candidates. If the mechanism selects a candidate which does not receive votes, the social cost is at least $\frac{n}{w+1}\cdot d_{\max}+\frac{n}{w+1}\cdot d_{\min}$. Therefore, the social cost achieved by the mechanism is lower bounded by 
\begin{align*}
&\frac{p_t}{w+1}\cdot\left((w-1)d_{\max}+2d_{min}\right)\cdot\frac{n}{w+1}+(1-p_t)\cdot \frac{n}{w+1}\cdot (d_{\max}+d_{\min})\\
=&\frac{p_t\cdot n}{w+1}\left(-\frac2{w+1}d_{\max}-\frac{w-1}{w+1}d_{\min}\right)+\frac{n}{w+1}(d_{\max}+d_{\min})\\
\ge &\frac{n}{w+1}\left(-\frac2{w+1}d_{\max}-\frac{w-1}{w+1}d_{\min}\right)+\frac{n}{w+1}(d_{\max}+d_{\min})\\
= &\frac{n}{w+1}\left(\frac{w-1}{w+1}d_{\max}+\frac2{w+1}d_{\min}\right).
\end{align*}

We combine it with the optimal social cost. Thus, the distortion is lower bounded by
\begin{align*}
&\frac{\frac{n}{w+1}\left(\frac{w-1}{w+1}d_{\max}+\frac2{w+1}d_{\min}\right)}{\frac{n}{w+1}\cdot d_{\min}}\\
=&\frac{(w-1)\sigma+2}{w+1}\\
=&1+\frac{w-1}{w+1}(\sigma-1).\qedhere
\end{align*}
\end{proof}

We now study the upper bound on the distortion.
We do this by analysing the SNTV mechanism, which achieves a distortion of at most $1+2\sigma$. 
\begin{mechanism}[SNTV Mechanism]
Given an election $\Gamma=(\Omega,M,\mathbf{a},w)$, then SNTV outputs a committee $C$ such that $C=\arg\max_{C'\in \mathcal{C}_w}n_{C'}$ where $n_{C'}=\sum_{y_j\in C'}n_j$. 
If there are multiple committees getting the most votes, then the mechanism breaks ties arbitrarily.
\end{mechanism}

To study the distortion of SNTV mechanism, we introduce the following lemma by assuming that $C^*$ is a committee in $\arg\min_{C\in \mathcal{C}_w} SC(\mathbf{x},C)$.

\begin{lemma}\label{lemma:UB}
The distortion of any anonymous mechanism is at most $$1+\frac{2\sigma}{n-n_{C^*}}\sum_{C\neq C^*} p_C(n_1,...,n_m)(n-n_C).$$
\end{lemma}

Our main proof tool is the triangle inequality. We defer proof details to the appendix.

Then, we notice that, by definition, $C=\arg\max_{C'\in \mathcal{C}_w}n_{C'}$, thus $n-n_C\le n-n_{C^*}$.
Lastly, owing to Lemma~\ref{lemma:UB}, we have

\begin{align*}
1+\frac{2\sigma}{n-n_{C^*}}\sum_{C\neq C^*} p_C(n_1,...,n_m)(n-n_C)\le1+2\sigma.
\end{align*}

\begin{theorem}
The distortion of SNTV mechanism is at most $1+2\sigma$.
\end{theorem}

%% file: sp.tex
\section{Distortion With Strategyproofness}
\label{sec:sp}
In this section, we focus our attention to strategyproof mechanisms.
Our study hinges upon the fact that on suitable elections, any strategyproof mechanism is \textit{independent of irrelevant candidates} (IIC).

\begin{definition}
A mechanism is \textit{independent of irrelevant candidates} (IIC) if the probability that the mechanism outputs a committee C is independent of the distribution of a fixed number of votes among candidates in $M\backslash C$ receive. Formally, for any committee $C=\{y_{l_1},\dots,y_{l_w}\} \subset M$, a mechanism is IIC if $p_C(\mathbf a) = p_C(n_{l_1},\dots,n_{l_w})$. 
\end{definition}

%
%

In order to prove that every strategyproof mechanism is IIC, we need to introduce the following candidate profile.

\begin{definition}[Candidate Profile II]
    Let us consider $m$ candidates in an $(m-1)$-dimensional Euclidean space. 
    Given $r\in\mathbb{R}$, we denote their locations with $y_1=(1,0,\dots,0),\dots, y_{m-1}=(0,0,\dots,1)$, and $y_m=(r,r,\dots,r)$.
    When $r$ is large enough, $d_{\min}$ is the distance between any two of the first $m-1$ candidates, and $d_{\max}$ is the distance between $y_m$ and any of the first $m-1$ candidates.
\end{definition}


We are then ready to state the following theorem.

\begin{theorem}
\label{lemma_independent}
In any election instance where candidates' positions are represented by Candidate Profile II, if $w<m-1$, any strategyproof mechanism must be IIC.
\end{theorem}

The proof idea is that we construct a subspace to analyze the distances between the voter and the candidates, and the expected cost of the voter is expressed in terms of these distances. By leveraging the strategyproofness of the mechanism, we show that two probabilities must be equal regardless of whether the voter chooses $y_i$ or $y_j$, thereby ensuring the mechanism satisfies IIC. 

\begin{proof}
For any voter $k$ and fixed actions of other voters $\mathbf{a}_{-k}$, we consider voter $k$'s two different actions and the corresponding action profiles $\mathbf{a}^1=(a_k=y_i,\mathbf{a}_{-k})$ and $\mathbf{a}^2=(a_k=y_j,\mathbf{a}_{-k})$. Denote $L=\{l_1,...,l_w\} \subset M$ a subset of candidates. We will show that for any committee $C=\{y_{l_1},...,y_{l_w}\}$ such that $y_i,y_j \notin C$, a strategyproof mechanism must have 
$p_C(\mathbf{a}^1)=p_C(\mathbf{a}^2)$. 
We then generalize this result to prove the theorem. There are two cases based on whether candidate $y_m$ is one of the candidates $y_i$ and $y_j$.

\textbf{Case 1, if $y_i = y_m$ or $y_j = y_m$}. Without loss of generality, we assume $y_j = y_m$.
Let $\alpha_1$ and $\alpha_2$ be real numbers such that $r/2\leq \alpha_1 \leq \alpha_2\leq (r+1)/2$. We define a subspace $U_L(\alpha_1,\alpha_2)$ as follows.
\begin{align*}
  \Bigl\{&(t_1,t_2,...,t_{m-1})\in \mathbb{R}^{m-1} | 
t_{l_1}=\frac{(m-2)r}{2}-(w-1)\alpha_1 -(m-w-2)\alpha_2, \\
& t_{l_2}=...=t_{l_w}=\alpha_1, t_i=\frac{r+1}{2}, t_h=\alpha_2, \forall h\notin L\cup \{i\} \Bigr\}.
\end{align*}
The construction of the subspace $U_L(\alpha_1,\alpha_2)$ has a twofold effect. First, the distances between voter $x_k\in U_L(\alpha_1,\alpha_2)$ to candidates $y_i$ and $y_j$ are the same. Second, the cost $cost(x_i,C)$ falls into three categories for all committees. This effect will facilitate us to represent voter $x_k$'s expected cost.

In particular, for any $x_k\in U_L(\alpha_1,\alpha_2)$, we have $d(x_k,y_i) = d(x_k,y_j)$. The two distances can be written as
\begin{align*}
     \biggl( \Bigl( \frac{r-1}{2} \Bigr)^2+t_{l_1}^2+(w-1)\alpha_1^2+(m-w-2)\alpha_2^2 \biggr)^{\frac12}. 
\end{align*}

Let $\eta:=d(x_k,y_i)^2+r+1$. For simplicity, we express the distances between $x_k$ and other candidates in terms of $\eta$. By simple calculation, we have that
\begin{align*}
&d(x_k,y_h)=\sqrt{\eta-2\alpha_2}, \,\,\,\,\,\ d(x_k,y_{l_1}) =\sqrt{\eta-2t_{l_1}} \\
&d(x_k,y_{l_2})=...=d(x_k,y_{l_w})=\sqrt{\eta-2\alpha_1}. 
\end{align*}
Since $r/2\le \alpha_1\le \alpha_2\le (r+1)/2$ and $h\notin L\cup \{i\}$, it is easy to check that $d(x_k,y_i)=d(x_k,y_j)\leq d(x_k,y_h)\leq d(x_k,y_{l_2})=...=d(x_k,y_{l_w})\leq d(x_k,y_{l_1})$. 
Next, we consider voter $k$'s cost. The distance from voter $k$ to the nearest candidate has three possibilities: $d(x_k,y_i)$, $d(x_k,y_{l_2})$ and $d(x_k,y_h)$ for different committees. 

Let $C'$ be a committee different from $C$ such that $y_i,y_j\notin C'$. Then, the summation $\sum_{C': y_i,y_j\notin C'}p_{C'}(\mathbf{a}^1)$ is the probability that the mechanism outputs one of these committees $C'$ when the action of voter $x_k$ is $\mathbf{a}^1$. Let $e_1(\mathbf{a}^1) = \sum_{C': y_i,y_j\notin C'}p_{C'}(\mathbf{a}^1)-p_C(\mathbf{a}^1)$ and $e_2(\mathbf{a}^1) = 1-\sum_{C': y_i,y_j\notin C'}p_{C'}(\mathbf{a}^1)$. 
Thus, when $a_k=y_i$, we can write the expected cost of voter $k$ as $\mathbb{E}[\cost(x_k)]= p_C(\mathbf{a}^1) d(x_k,y_{l_2})+e_1(\mathbf{a}^1) d(x_k,y_h)+e_2(\mathbf{a}^1) d(x_k,y_i)$.

Similarly, when $a_k=y_j$, we have that $\mathbb{E}[\cost(x_k)]=p_C(\mathbf{a}^2) d(x_k,y_{l_2})+e_1(\mathbf{a}^2) d(x_k,y_h)+e_2(\mathbf{a}^2) d(x_k,y_i)$.

Since the mechanism is strategyproof, voter $x_k$ should derive the same cost when she votes for $y_i$ or $y_j$. 
We then have that 
\begin{align}\label{eq:square}
    &\left(p_C(\mathbf{a}^1)-p_C(\mathbf{a}^2)\right)  d(x_k,y_{l_2})+\left(e_1(\mathbf{a}^1)-e_1(\mathbf{a}^2)\right) d(x_k,y_h)\nonumber\\
    =&\left(e_2(\mathbf{a}^2)-e_2(\mathbf{a}^1)\right) d(x_k,y_i).
\end{align}

We square both sides of the equation~\eqref{eq:square}. The terms $d^2(x_k,y_{l_2})$, $d^2(x_k,y_h)$ and $d^2(x_k,y_i)$ do not involve a square root. However, the term $d(x_k,y_{l_2})\cdot d(x_k,y_h)$ still contains a square root. Since we are considering the Euclidean distance, and the equation should hold for any $\alpha_1$ and $\alpha_2$ such that $r/2\leq \alpha_1 \leq \alpha_2\leq (r+1)/2$. Therefore, the coefficient of the term $d(x_k,y_{l_2})\cdot d(x_k,y_h)$ must be 0. That is,  
\[2\cdot\left(p_C(\mathbf{a}^1)-p_C(\mathbf{a}^2)\right)\cdot \left(e_1(\mathbf{a}^1)-e_2(\mathbf{a}^2)\right)=0.\] 
Consequently, we conclude that $p_{C}(\mathbf{a}^1)=p_{C}(\mathbf{a}^2),e_1(\mathbf{a}^1)=e_1(\mathbf{a}^2)$ and $e_2(\mathbf{a}^1)=e_2(\mathbf{a}^2)$.

\textbf{Case 2, if $y_i\neq y_m$ and $y_j\neq y_m$.} The proof follows a similar approach as in Case 1, but the process of constructing the subspace to establish these equations is more intricate. To keep our primary discussion on track, we defer the detailed construction of the subspace for this case to the Appendix.

These two cases illustrate that, irrespective of the committee $C$, when a single voter's decision leads to the election of a candidate from $M\backslash C$, it does not impact the probability of the mechanism producing committee $C$. By recursively applying this proof, we can establish that the same outcome holds true even if multiple voters alter their choices, resulting in the election of candidates from $M\backslash C$. In other words, the mechanism remains unaffected by candidates who are irrelevant to the final result.
\end{proof}

For any committee $C=\{y_{l_1},\dots,y_{l_w}\} \subset M$, the probability that an IIC mechanism outputs $C$ depends only on the number of votes the candidates $y_{l_1},\dots,y_{l_w}$ receive. 
%
%
Note that given any strategyproof mechanism $f'$, there always exists a randomized, anonymous strategyproof mechanism $f$ whose distortion is not worse than $f'$. Specifically, $f$ can be obtained by applying a uniformly chosen random permutation to the set of voters before applying $f'$. 
So, without loss of generality, we consider randomized, anonymous strategyproof mechanisms. Because of Theorem~\ref{lemma_independent}, these mechanisms are IIC. 

Recall that the lower bound of distortion in Theorem~\ref{thm:lb_nonsp} can also be applied to the strategyproof setting. 
Moving forward,  we present an upper bound for randomized strategyproof mechanisms. 
%
%
En route to this result, we retrieve a sufficient condition for strategyproofness and define a mechanism that meets this sufficient condition.

\begin{definition}(Monotonicity)
An IIC mechanism is monotone if the probability of a committee $C=\{y_{l_1},\dots,y_{l_w}\}$ being elected weakly increases when the number of votes received by a candidate in $C$ weakly increases and others are unchanged. In particular, for a $w$-winner election, for any committee $C$ and any $y_{l_i} \in C$, it holds that $p_C(n_{l_i}+1,\mathbf{n_{-l_i}})\ge p_C(n_{l_i},\mathbf{n_{-l_i}})$,
where $\mathbf{n_{-l_i}}$ denotes the number of votes the candidates in $C$ excepts $y_{l_i}$ received.  
\end{definition}
\begin{theorem}\label{theorem_monotone}
Any monotone mechanism is strategyproof.
\end{theorem}

Next, we present an upper bound of distortion. Therefore, we need to study a strategyproof mechanism. Recalling the SNTV mechanism, it is not strategyproof.
\begin{proposition}\label{prop:notsp}
The SNTV is not a strategyproof mechanism.
\end{proposition}

\begin{proof}
Suppose there are four candidates: $y_1, y_2, y_3$ and $y_4$. The goal is to elect 2 winners. There are 7 voters. We have $y_1, y_2$ and $y_3$ receive $2$ votes respectively. We consider the last voter. She locates at $y_4$. As for the distances, we assume that $d(y_1, y_4)>d(y_2, y_4)> d(y_3, y_4)$. If the last voter reports honestly, that is $y_4$, the candidate pair $\{y_1,y_2\}$, $\{y_1,y_3\}$ and $\{y_2, y_3\}$ should be elected arbitrarily by SNTV mechanism. If she misreports to $y_3$, then $y_3$ gets 3 votes. Thus, either $\{y_1,y_3\}$ or $\{y_2, y_3\}$ is elected by SNTV mechanism. The cost of the voter decreases. 
\end{proof}

%

To conclude, we introduce and study the Random Sequential Dictator mechanism.

\begin{mechanism}\label{mec_independent} (Random Sequential Dictator)
For a given election $\Gamma=(\Omega,M,\mathbf{a},w)$, the voters' actions are arranged in a sequence randomly. The mechanism always outputs the first $w$ different candidates as the committee.
\end{mechanism}

The strategyproofness of Random Sequential Dictator is apparent from its dictator nature. In the sequence of indexing the candidates, when it comes to voter $x_i$'s action, if the first $w$ candidates are not entirely determined yet, then voter $x_i$ should vote for the nearest candidate to her to minimize her cost; if the first $w$ candidates are already determined, then voter $x_i$'s action will not change the output committee of the mechanism. Then, in order to derive the distortion of the mechanism, we need the probability function of each committee so that we can use Lemma~\ref{lemma:UB}.

\begin{proposition}\label{prop:RSD}
When $w=2$, the probability function of Random Sequential Dictator for committee $C=\{y_k,y_l\}$ is \[p_C(n_k,n_l)=\frac{n_k}{n-n_l}+\frac{n_l}{n-n_k}-\frac{n_k+n_l}{n}.\] 
\end{proposition}

\begin{proof}
We consider the probability of ${y_k,y_l}$ being selected. For the case where $y_k$ is the first chosen candidate and $y_l$ is the second, the probability is $\frac{n_k}{n}\cdot\frac{n_l}{n-n_k}$. For the case where $y_l$ is the first chosen candidate and $y_k$ is the second, the probability is $\frac{n_l}{n}\cdot\frac{n_k}{n-n_l}$. Thus, we have
\begin{align*}
p_C(n_k,n_l)=&\frac{n_k}{n}\cdot\frac{n_l}{n-n_k}+\frac{n_l}{n}\cdot\frac{n_k}{n-n_l}\\
=&n_l\cdot\frac{n_k}{n(n-n_k)}+n_k\cdot\frac{n_l}{n(n-n_l)}\\
=&\frac{n_k}{n-n_l}+\frac{n_l}{n-n_k}-\frac{n_k+n_l}{n}.\qedhere
\end{align*}
\end{proof}

Lastly, we compute the distortion upper bound of the mechanism. 

\begin{theorem}\label{theorem:PI}
The distortion of Random Sequential Dictator is upper bounded by $1+4\sigma$ when $w=2$.
\end{theorem}

When $w>2$, it is difficult to analyze the performance of Random Sequential Dictator mechanism. Thus, we apply Lemma~\ref{lemma:UB} to get an upper bound. 

\begin{proposition}\label{prop:rsd}
    The distortion of Random Sequential Dictator is upper bounded by $1+2(n-w)\sigma$ when $w>2$.
\end{proposition}

Random Dictator is a special case of Random Sequential Dictator when $w=1$. For every candidate $y_i\in M$, the winning probability is $n_i/n$. \cite{DBLP:conf/sigecom/FeldmanFG16} showed that Random Dictator mechanism achieves a distortion of exactly 3.  By leveraging on Theorem~\ref{lemma_independent}, we show that Random Dictator is the only anonymous strategyproof mechanism that has finite distortion. 

\begin{theorem}\label{theorem:RD}
In single-winner elections, the unique anonymous strategyproof mechanism using single-candidate ballots with finite distortion is Random Dictator. 
\end{theorem}


%
To conclude, we consider deterministic mechanisms.
In particular, we show that no anonymous deterministic strategyproof mechanism can achieve  finite distortion.
Lastly, we show that anonymousness plays a crucial role into showing the unboundness of the lower bound by considering the Sequential Dictator mechanism and investigating its distortion. 

\begin{theorem}\label{theorem:multi_deter}
For a $w$-winner election, no anonymous deterministic strategyproof mechanism can achieve finite distortion. 
\end{theorem}

\begin{mechanism} (Sequential Dictator)
Given an election $\Gamma=(\Omega,M,\mathbf{a},w)$, the voters cast their actions by a predetermined order. The mechanism always outputs the first $w$ different candidates as the committee.
\end{mechanism}

\begin{theorem}\label{theorem:SD}
Sequential Dictator is a deterministic strategyproof mechanism and its distortion is at most $2(n-w)\sigma+1$.
\end{theorem}

%% file: appendix.tex
\section{Missing proofs in Section \ref{sec:non-sp}}
\begin{proof}[Proof of Lemma \ref{lemma:UB}]
First, the distortion $dist(f,\Gamma)$ can be represented by
\begin{align*}
\sup_{\mathbf{x}\in \chi(\Gamma)} \frac{\sum_{C\in \mathcal{C}_w} p_C(n_1,...,n_m)\SC(\mathbf{x}, C)}{\SC(\mathbf{x}, C^*)}.
\end{align*}
For any voter $i$, let $y_k=\arg\min_{y\in C} d(x_i,y)$ and $y_k^*=\arg\min_{y^*\in C^*} d(x_i,y^*)$. Denote $d_i(C,C^*) = d(y_k,y_k^*)$. On the one hand, for any $C\neq C^*$, note that the social cost $\SC(\mathbf{x},C)$ can be upper bounded by
\begin{align*}
\SC(\mathbf{x},C)=&\sum_{i:a_i\in C}d(x_i,C)+\sum_{i:a_i\notin C}d(x_i,C) \\
\leq & \,\,\ \sum_{i:a_i\in C}d(x_i,C^*)+\sum_{i:a_i\notin C}d(x_i,C)\\
= & \,\,\ \SC(\mathbf{x},C^*) + \sum_{i:a_i\notin C}(d(x_i,C)-d(x_i,C^*))\\
\leq & \,\,\ \SC(\mathbf{x},C^*) + \sum_{i:a_i\notin C}d_i(C, C^*)\\
\leq & \,\,\ \SC(\mathbf{x},C^*) + (n-n_C)d_{\max}.
\end{align*}

The first inequality arises because the voters choose candidates from $C$ as they are closer to $C$ than to candidates in $C^*$. The second inequality is due to triangle inequality. The third inequality is because the distance between any two candidates is upper bounded by $d_{\max}$. 
Summing up all possible committees on both sides of inequality, we obtain 
\[\sum_{C\in \mathcal{C}_w} p_C(n_1,...,n_m)\SC(\mathbf{x}, C) 
\le  \SC(\mathbf{x}, C^*)+\sum_{C\neq C^*} p_C(n_1,...,n_m)(n-n_C)d_{\max}.\]
On the other hand, we have that 
\begin{align*}
   \SC(\mathbf{x},C^*) 
   &=\frac12 \sum_{i\in N}2d(x_i,C^*)\\
   &\geq \frac12 \sum_{i\in N}(d(x_i,C^*)+d(x_i,a_i))\nonumber\\
   &\geq \frac12 \sum_{i\in N}d(a_i,C^*)\\
   &=\frac12\sum_{i:a_i\in C^*}d(a_i,C^*)+\frac12\sum_{i:a_i\notin C^*}d(a_i,C^*)\\
    &\geq \frac12 (n-n_{C^*})d_{\min}.
\end{align*}
The first inequality holds because $a_i$ is the nearest candidate to location $x_i$. 
The second inequality is due to the triangle inequality, and the last inequality holds because for voters who vote for the candidate in $C^*$, the distance $d(a_i, C^*)$ is 0 and for voters who vote for the candidate not in $C^*$, the distance $d(a_i, C^*)$ is lower bounded by $d_{\min}$.

Consequently, the distortion is upper bounded by 
\begin{align*}
& \frac{\SC(\mathbf{x}, C^*)+\sum_{C\neq C^*} p_C(n_1,...,n_m)(n-n_C)d_{\max}}{\SC(\mathbf{x}, C^*)} \\
\le& 1+\frac{\sum_{C\neq C^*} p_C(n_1,...,n_m)(n-n_C)d_{\max}}{\frac12 (n-n_{C^*})d_{\min}} \\
=& 1+\frac{2\sigma}{n-n_{C^*}}\sum_{C\neq C^*} p_C(n_1,...,n_m)(n-n_C).\qedhere
\end{align*}
\end{proof}

\section{Missing proofs in Section \ref{sec:sp}}
\begin{proof}[Missing proof of Theorem \ref{lemma_independent}]
\textbf{Case 2, if $y_i\neq y_m$ and $y_j\neq y_m$.} 
Depending on whether the candidate $y_m$ is included in the committee $C=\{y_{l_1},\dots,y_{l_w}\}$, we consider two subcases and construct a subspace for each case correspondingly. \\
{\bf Subcase 2.1}, $y_m\in C$.\\
Without loss of generality, we assume that $y_{l_1}=y_m$. For any $\alpha_1$ and $\alpha_2$ such that $0\leq \alpha_1 \leq \alpha_2\leq 1/2$, we define a subspace $U_{L}(\alpha_1,\alpha_2)$ as
\begin{align*}
\Bigl\{&(t_1,t_2,...,t_{m-1})\in \mathbb{R}^{m-1}|t_i=\frac12, t_j=\frac12, 
 t_{l_2}=...=t_{l_w}=\alpha_1,\\
 &t_h=\alpha_2,\forall h\notin L \cup\{i,j\}\Bigr\}.
\end{align*}

Note that $d(x_k,y_i)=d(x_k,y_j)\leq d(x_k,y_h)\leq d(x_k,y_{l_2})=\cdots=d(x_k,y_{l_w})\leq d(x_k,y_{l_1}), \forall h\notin L\cup \{i,j\}$. 
We consider three distances $d(x_k,y_i), d(x_k,y_{l_2}), d(x_k,y_h)$ and their corresponding possibilities. By a similar argument as Case 1, we can derive that $p_{C}(\mathbf{a}^1)=p_{C}(\mathbf{a}^2)$.

{\bf Subcase 2.2}, $y_m\notin C$.\\
For any $\alpha_1, \alpha_2$ and $\alpha_3$ such that $0\leq\alpha_1 \leq \alpha_2\leq \alpha_3 \leq 1/2$, we define a subspace $U_{L}(\alpha_1,\alpha_2,\alpha_3)$ as

\begin{align*}
\Bigl\{&(t_1,t_2,...,t_{m-1})\in \mathbb{R}^{m-1}|t_i=\frac12, t_j=\frac12, t_{l_1}=\alpha_1, 
 t_{l_2}=...=t_{l_w}=\alpha_2, \\
 &t_h=\alpha_3,\forall h\notin L_2\cup\{i,j\}\Bigr\}.
\end{align*}
Note that $d(x_k,y_i)=d(x_k,y_j)\leq d(x_k,y_h)\leq d(x_k,y_{l_2})=...=d(x_k,y_{l_w})\leq d(x_k,y_{l_1})\leq d(x_k,y_m), \forall h\notin L\cup \{i,j\}$. We also consider the three distances $d(x_k,y_i),d(x_k,y_{l_2})$ and $d(x_k,y_h)$. Similarly, we have 

\[p_C(\mathbf{a}^1)+\sum_{C':y_m\in C',C'\backslash\{y_m\}\subset C}p_{C'}(\mathbf{a}^1)
=p_C(\mathbf{a}^2)+\sum_{C':y_m\in C',C'\backslash\{y_m\}\subset C}p_{C'}(\mathbf{a}^2).\]

From the Subcase 2.1, we know that $p_{C'}(\mathbf{a}^1)=p_{C'}(\mathbf{a}^2)$ for any $C'$ such that $y_m\in C',C'\backslash\{y_m\}\subset C$. So we can get $p_C(\mathbf{a}^1)=p_C(\mathbf{a}^2)$.   
\end{proof}

\begin{proof}[Proof of Theorem~\ref{theorem_monotone}]
For any monotone mechanism, consider a location profile $\mathbf{x}$ and an arbitrary voter $x_i$. Suppose that $y_{k^*}$ is the nearest candidate to voter $x_i$ so that in a strategyproof mechanism, $x_i$'s consistent action is voting for $y_{k^*}$. To prove the theorem, We compare $x_i$'s cost when she votes for $y_{k^*}$ and when she inconsistently votes for some other candidate $y_{k'}$.

We partition all possible committees into three mutually exclusive and collectively exhaustive sets, depending on whether a committee contains $y_{k^*}$ and $y_{k'}$. That is, the set of all possible committees $\mathcal{C}_w = \mathcal{C}_{(1)} \cup \mathcal{C}_{(2)} \cup \mathcal{C}_{(3)}$, where $\mathcal{C}_{(1)} = \{C ~|~ y_{k^*}\notin C, y_{k'} \notin C \}$, $\mathcal{C}_{(2)} = \{C ~|~ y_{k^*}\notin C, y_{k'} \in C \}$, and $\mathcal{C}_{(3)} = \{C ~|~ y_{k^*}\in C \}$.

For any committee $C \in \mathcal{C}_{(1)}$, since the mechanism is IIC, $p_C(a_i=y_{k^*},\mathbf{a}_{-i}) = p_C(a_i=y_{k'},\mathbf{a}_{-i})$. Therefore, voter $x_i$'s expected cost, when the mechanism outputs a committee $C$ that falls in this category, remains the same no matter if her action is either $a_i=y_{k^*}$ or $a_i=y_{k'}$. Denote $Q_1:=\sum_{C \in \mathcal{C}_{(1)}} p_C$. 

For any committee $C \in \mathcal{C}_{(2)}$, voter $x_i$'s expected cost, when she takes action $a_i=y_{k^*}$, is $\sum_{C \in \mathcal{C}_{(2)}} p_C(a_i=y_{k^*}, \mathbf{a}_i) d(x_i,C) = \sum_{C \in \mathcal{C}_{(2)}} p_C(n_{k'}, \mathbf{n_{-k'}}) d(x_i,C)$, where $\mathbf{n_{-k'}}$ is the number of votes other candidates in $C$ receive. Her expected cost, when she takes action $a_i=y_{k'}$, is $\sum_{C \in \mathcal{C}_{(2)}} p_C(a_i=y_{k'}, \mathbf{a}_i) d(x_i,C) = \sum_{C \in \mathcal{C}_{(2)}} p_C(n_{k'}+1, \mathbf{n_{-k'}}) d(x_i,C)$. Since the mechanism is monotone, we know 
\begin{align*}
 \sum\limits_{C \in \mathcal{C}_{(2)}} \bigl( p_C(n_{k'}+1, \mathbf{n_{-k'}}) - p_C(n_{k'}, \mathbf{n_{-k'}}) \bigr) d(x_i,C)  \ge   0.
\end{align*}

Denote $Q_2:=\sum_{C \in \mathcal{C}_{(2)}} p_C(n_{k'}+1, \mathbf{n_{-k'}})$. For any committee $C \in \mathcal{C}_{(3)}$, we know that the total probability that the mechanism outputs a committee that falls into this set is $1 - Q_1 - Q_2$. So, voter $x_i$'s expected cost, when she takes action $a_i=y_{k'}$, is 
\begin{align*}
    (1 - Q_1 - Q_2) d(x_i, C) = (1 - Q_1 - Q_2) d(x_i, y_{k^*}).
\end{align*}
While voter $x_i$'s expected cost, when she takes action $a_i=y_{k^*}$, is 
\[(1 - Q_1 - \sum_{C \in \mathcal{C}_{(2)}} p_C(n_{k'}, \mathbf{n_{-k'}}) ) d(x_i, C) 
    = (1 - Q_1 - \sum_{C \in \mathcal{C}_{(2)}} p_C(n_{k'}, \mathbf{n_{-k'}}) ) d(x_i, y_{k^*}).\]
Hence, the expected cost of voter $x_i$ in this category is reduced by 
\begin{align*}
   \textstyle \sum_{C \in \mathcal{C}_{(2)}} \bigl( p_C(n_{k'}+1, \mathbf{n_{-k'}}) - p_C(n_{k'}, \mathbf{n_{-k'}}) \bigr) d(x_i,y_{k^*}) \textstyle \ge 0.
\end{align*}

Note that $y_{k^*}$ is the nearest candidate to voter $i$. Hence, we have $d(x_i,C)\ge d(x_i, y_{k^*})$, where $C \in \mathcal{C}_{(2)}$. Therefore, 
$\cost(x_i,f(a_i=y_{k'},\mathbf{a}_{-i}))\ge \cost(x_i, f(a_i=y_{k^*},\mathbf{a}_{-i}))$, which concludes that the mechanism is strategyproof.
\end{proof}

\begin{proof}[Proof of Theorem~\ref{theorem:PI}]
We suppose that the optimal candidate pair is $\{y_{k^*},y_{l^*}\}$. First, using Lemma~\ref{lemma:UB}, we have that the distortion is upper bounded by
\begin{align}\label{eq:rsd}
&1+\frac{2\sigma}{n-n_{C^*}}\sum_{C\neq C^*} p_C(n_1,...,n_m)(n-n_C)\nonumber\\
=&1+\frac{2\sigma}{n-n_{C^*}}\left(\sum_{C\in \mathcal{C}_w} p_C(n_1,...,n_m)(n-n_C)-p_{C^*}(n_1,..,n_m)(n-n_{C^*})\right).
\end{align} 
Next, we plug the probability function in Proposition~\ref{prop:RSD} into the summation. Thus, we have
\begin{align*}
    &\sum_{\{y_k,y_l\}\in \mathcal{C}_2} p_C(n_k,n_l)(n-n_C) \\
   =& \sum_{\{y_k,y_l\}\in \mathcal{C}_2}\left(\frac{n_k}{n-n_l}+\frac{n_l}{n-n_k}-\frac{n_k+n_l}{n}\right)(n-n_k-n_l) \\
   =&\sum_{\{y_k,y_l\}\in \mathcal{C}_2}\left(\frac{(n_k+n_l)^2}{n}-\frac{n_k^2}{n-n_l}-\frac{n_l^2}{n-n_k}\right) \\
   =& \frac{(\sum_{k=1}^m n_k)^2}{n}+\sum_{k=1}^m\left(\frac{(m-2)n_k^2}{n}-\sum_{l\neq k}\frac{n_k^2}{n-n_l}\right). 
\end{align*}
For any $k\in\{1,\cdots,m\}$, we have $\frac{n_k^2}{n}-\frac{n_k^2}{n-n_l}\leq 0$. It implies that $\frac{(m-2)n_k^2}{n}-\sum_{l\neq k}\frac{n_k^2}{n-n_l}\leq 0$. We remain two terms containing $n_{k^*}$ and $n_{l^*}$, and get
 \[\sum_{\{y_k,y_l\}\in \mathcal{C}_2} p_C(n_k,n_l)(n-n_C)\leq n-\frac{n_{k^*}^2}{n-n_{l^*}}-\frac{n_{l^*}^2}{n-n_{k^*}}.\]

We plug it into \eqref{eq:rsd} and prove its distortion. So, we further get $\dist(f,\Gamma)$ is upper bounded by
\begin{align*}
     &1+\frac{2\sigma}{n-n_{C^*}}\big(n-\frac{n_{C^*}^2}n\big)=1+2\sigma\cdot\frac{n+n_{C^*}}n\le1+4\sigma.
      \qedhere
\end{align*}
\end{proof}

\begin{proof}[Proof of Proposition~\ref{prop:rsd}]
By Lemma~\ref{lemma:UB}, we have
\[\dist(f,\Gamma) \le 1+\frac{2\sigma}{n-n_{C^*}}\sum_{C\in\mathcal{C}_w} p_C(n_{l_1},...,n_{l_w})(n-n_C).\]
When $w> 2$, we can easily get 
\begin{align*}
&1+\frac{2\sigma}{n-n_{C^*}}\sum_{C\in\mathcal{C}_w} p_C(n_{l_1},...,n_{l_w})(n-n_C)\\
\le&1+2\sigma\cdot\frac{n-w}{n-n_{C^*}}\\
\le& 1+2(n-w)\sigma.\qedhere
\end{align*}
\end{proof}

\begin{proof}[Proof of Theorem~\ref{theorem:RD}]
Using Theorem~\ref{lemma_independent}, let $p_k(n_k)$ denote the probability of electing $y_k$, where $n_k$ is the number of votes received by $y_k$. We fix a candidate $y_i$ and choose another candidate $y_k$ arbitrarily. For all $y_k\neq y_i$ and $n_k\in[n-1]$, we have $p_i(1)+p_k(n_k)=p_k(n_k+1)+p_i(0)$. Therefore, we know that $p_k(n_k+1)-p_k(n_k)=p_i(1)-p_1(0)$. Notice that $p_i(1)-p_1(0)$ is a constant. This implies that for all $k\neq i$, the function $p_k$ is a linear function. Moreover, all these functions share the same coefficient $p_i(1)-p_1(0)$. Because the sum of probabilities is always 1, the function $p_i$ must also be linear with the same coefficient. We use $c$ to denote the coefficient and suppose that $p_k(n_k)=c\cdot n_k+b_k(b_k\geq 0)$. Consider Candidate Profile II and $n$ voters are located exactly at $y_m$. For this location profile $\mathbf{x^1}$, we have $OPT(\mathbf{x^1})=0$. The expected social cost is

\[\sum_{k\neq m}p_k(0)\cdot n\cdot d_{\max} = n \cdot d_{\max} \sum_{k\neq m}b_k.\]

If the mechanism has finite distortion, we conclude that $b_k=0$ for all $k\neq m$. Then, we assume that $n$ voters are located at $y_1$. For this location profile $\mathbf{x^2}$, we have $OPT(\mathbf{x^2})=0$. The expected social cost is

\[\sum_{k\neq 1,m}p_k(0)\cdot n\cdot d_{\min}+ p_m(0)\cdot n\cdot d_{\max}= n \cdot d_{\max} \cdot b_m.\]

Thus, we find that $b_m=0$. Since the sum of the probability $\sum_{k= 1}^m p_k(n_k)$ is 1, we conclude that the coefficient $c$ is $1/n$. To sum up, the only anonymous strategyproof mechanism with finite distortion for single-winner election is Random Dictator.
\end{proof}

\begin{proof}[Proof of Theorem~\ref{theorem:multi_deter}]
For a deterministic mechanism, it outputs a committee with probability 1. According to Theorem~\ref{lemma_independent}, this means that $p_C$ should be 0 or 1 for all $C\in \mathcal{C}_w$. We reconsider Candidate Profile II and set $m=w+2$. So any winning committee is obtained by excluding two candidates. 

First, we denote the committee that excludes candidates $y_i, y_j\in M$ by $C_{i,j}:=M\backslash\{y_i, y_j\}$ for simplicity. W.l.o.g., we only need to consider anonymous mechanisms. We prove that for any committee $C$ that contains $y_{w+2}$, when the candidate $y_{w+2}$ does not receive any vote and everyone else in $C$ receive non-zero votes, and candidates in $C$ receive less than $n$ votes, then the probability of such committee is elected is 0. Let us consider a voters' location profile $\mathbf{x}^1$ in which $h_k$ voters' location is at $y_k$, where $h_k \ge 1$ for $k=1,\dots, w$ and $ \sum_{k=1}^{w} h_k=n$. In this instance, the unique optimal committee is $C_{w+1, w+2}$ and $\OPT(\mathbf{x}^1)=0$. 
Therefore, for any strategyproof mechanism to achieve a finite distortion, it should output the committee $C_{w+1, w+2}$ deterministically. Hence, for any committee $C$ which can be an output of a randomized strategyproof mechanism, if the committee $C$ includes $y_{w+2}$, the probability that the mechanism outputs the committee $C$ is 0. 
By Theorem \ref{lemma_independent}, any strategyproof mechanism is independent of irrelevant candidates. 
Therefore, considering the committee $C_{i,w+1}$ for any $i\le w$, we have $p_{C_{i,w+1}}(h_1,\cdots,h_{i-1},h_{i+1},\dots,h_w,0)=0$ as long as $h_k\ge 1$ for $k\neq i$ and $\sum_{k=1}^{i-1} h_k +\sum_{k=i+1}^{w}h_k<n$. 
Actually, we can apply the same argument on any $w$ candidates among the first $w+1$ candidates. Therefore, we are able to claim that for any $i,j\le w+1$, $p_{C_{i,j}}(h_1,\dots,h_{i-1},h_{i+1},\dots,h_{j-1},h_{j+1},\cdots, h_{w+1},0)=0$ as long as $h_k\ge 1$ for $k\in [w+1]\backslash i\backslash j$ and $\sum_{k\in [w+1]\backslash i\backslash j}h_k<n$. It indicates that if the sum of votes received by each candidate in a committee is $w+2$, this committee should be elected deterministically. Furthermore, if there is a candidate in a committee who receives 0 vote, the probability that such a committee is elected should be 0.

Next, suppose that $w+2$ voters participate in the election. We consider the location profile where voter $i$ locates at $y_i$. Specifically, we have $\mathbf{x}^1=(x_1=y_1,...,x_{w+2}=y_{w+2})$. There are two cases depending on whether the mechanism elects $y_{w+2}$. However, we do not use $d_{\max}$ or $d_{\min}$ in the proof, that is, the two cases are the same. Therefore, we consider the case where the mechanism does not elect $y_{w+2}$ as an example.

If the mechanism does not elect $y_{w+2}$,  we assume, without loss of generality, that it outputs $C_{w+1,w+2}$. Then, we have
    \begin{align*}
        &p_{C_{w+1,w+2}}(1,1,...,1)=1;\\
        &p_C(1,1,...,1)=0, \forall C\neq C_{w+1,w+2}.
    \end{align*}
    Next, depending on whether $w$ is even or odd, we consider two different series of location profiles where $w+2$ voters are located at $w+1$ candidates. 
    
    If $w$ is even, we consider a series of location profiles where 2 voters are located at $y_2$. That is, there are $w+1$ location profiles in total. Taking $\mathbf{x}^2=(x_1=x_2=y_2, x_3=y_3,..., x_{w+2}=y_{w+2})$ as an example, we have 
    
    \[\sum_{j=3}^{w+2}p_{C_{1,j}}(2,1,...,1)=1.\] 
    
    Similarly, we get $w$ equations from the other $w$ location profiles. Note that there are always 2 voters at $y_2$, so we can omit the input of $p_C$ for simplicity. Summing the $w+1$ equations, we get $2\cdot\left(\sum_{j,k\neq 2} p_{C_{j,k}}\right)=w+1$ which contradicts the condition where $p_C$ should be 0 or 1.

If $w$ is odd, we consider another series of location profiles where 2 voters are located at $y_{w+2}$. There are also $w+1$ location profiles in total. For $\mathbf{x}^3=(x_1=y_1,...,x_w=y_w, x_{w+1}=x_{w+2}=y_{w+2})$, the mechanism should output $C_{w+1,w+2}$ deterministically. For the other $w$ location profiles, take $\mathbf{x}^4=(x_1=y_1,...,x_{w-1}=y_{w-1},x_w=y_{w+1},x_{w+1}=x_{w+2}=y_{w+2})$ as an example. We have

    \[\sum_{j=1}^{w-1}p_{C_{j,w}}(1,...,1,2)=1.\]
    
    Similarly, we get $w-1$ equations from the other $w-1$ location profiles. We can also omit the input of $p_C$ for simplicity. Summing the $w$ equations, we get $2\cdot\left(\sum_{j,k\neq w+1,w+2} p_{C_{j,k}}\right)=w$ which also contradicts the condition that $p_C$ should be 0 or 1. 

    Therefore, no anonymous deterministic strategyproof single-candidate ballots mechanism can achieve finite distortion.
\end{proof}

\begin{proof}[Proof of Theorem~\ref{theorem:SD}]
The proof of strategyproofness is similar to the Random Sequential Dictator mechanism.

To prove the distortion upper bound of sequential dictator, we use an argument similar to the proof of Theorem~\ref{theorem:PI}. Given a location profile $\mathbf{x}$ and suppose that the optimal committee is $C^*$. Denote $f(\mathbf{a})=C$ the output of sequential dictator. The social cost achieved by the mechanism is upper bounded by 

\[\SC(\mathbf{x}, C^*) + (n-n_C)d_{\max}.\]

The optimal social cost $SC(C^*,\mathbf{x})$ is lower bounded by 

\[\frac12 (n-n_{C^*})d_{\min}.\]

Consequently, the distortion is upper bounded by

\[1+2\sigma\frac{n-n_C}{n-n_{C^*}}.\]

To retrieve the upper bound, we set $n_C = w$ and $n_{C^*}=n-1$. Therefore, the distortion is at most $2(n-w)\sigma+1$.
\end{proof}